\documentclass[preprint,authoryear,11pt]{elsarticle}

\usepackage{epsfig}

\usepackage{amssymb}

\usepackage[ps2pdf,%
a4paper=true,%
breaklinks=true,%
colorlinks=true,%
pdfauthor={Emil Khalisi},%
pdftitle={Dust clouds and plasmoids in Saturn's magnetosphere...},%
pdfsubject={Advances in Space Research (2017)}
]{hyperref}

\journal{Advances in Space Research}

\usepackage{rotating}
\usepackage[british]{babel}   
\usepackage{colortbl}
\definecolor{grey10}{RGB}{230,230,230}

\begin{document}

\begin{frontmatter}



\title{Dust clouds and plasmoids in Saturn's Magnetosphere\\ as seen with four {\it Cassini} instruments}


\author{Emil Khalisi\fnref{myfootnote}}
\fntext[myfootnote]{Corresponding author: ekhalisi@khalisi.com}
\address{Max-Planck-Institute for Nuclear Physics, Saupfercheckweg 1, D--69117 Heidelberg, Germany}
%

 \ead[url]{DOI: http://dx.doi.org/10.1016/j.asr.2016.12.030}



\begin{abstract}

We revisit the evidence for a "dust cloud" observed
by the {\it Cassini} spacecraft at Saturn in 2006.
The data of four instruments are simultaneously compared
to interpret the signatures of a coherent swarm of
dust that would have remained near the equatorial plane
for as long as six weeks.
The conspicuous pattern, as seen in the dust counters of the
Cosmic Dust Analyser (CDA), clearly repeats on three
consecutive revolutions of the spacecraft.
That particular cloud is estimated to about 1.36 Saturnian
radii in size, and probably broadening.
We also present a reconnection event from the magnetic field
data (MAG) that leave behind several plasmoids like those
reported from the Voyager flybys in the early 1980s.
That magnetic bubbles happened at the dawn side of Saturn's
magnetosphere.
At their nascency, the magnetic field showed a switchover of
its alignment, disruption of flux tubes and a recovery on
a time scale of about 30 days.
However, we cannot rule out that different events might have
taken place.
Empirical evidence is shown at another occasion when a
plasmoid was carrying a cloud of tiny dust particles such
that a connection between plasmoids and coherent dust clouds
is probable.\\
\end{abstract}


\end{frontmatter}

\parindent=0.5 cm

\section{Introduction}

Interplanetary dust clouds are local density enhancements
of particles of a specific mass type.
They would usually be relicts of dissolved comets,
debris of an asteroidal collision,
ejecta from planets or moons,
and a few may also go back to jet streams from active
bodies,
or coronal mass ejections.
In the vast range of patterns, the characteristics of dust
will vary on all dimensions: size, density, mass,
lifetime, and more (see, e.g., \cite{gruen-etal_2004}).

During the {\it Pioneer} and both {\it Voyager} flybys at
Saturn in the 1980s,
regions of plasma material were found on the day-side of
the Saturnian magnetosphere (\cite{bridge-etal_1981},
\cite{bridge-etal_1982}).
These are local spots of relatively dense and cold plasma
at distances beyond 16 Saturnian radii
(1 $R_{\rm S}$ = 60,268 km).
In particular, {\it Voyager 2}'s ion spectrometer suggested
that numerous of such isolated ``plasma blobs'' float
closely to the Saturnian magnetopause.
Dust grains, which would be very much more massive than
plasma components, were not reported.
We accidentally discovered in the {\it Cassini} data
one striking event that showed a local enhancement of dust
particles resembling a ``dust blob'' remaining for a few
revolutions of the spacecraft (\cite{khalisi-etal_2015}).
Now, we will investigate that event in more detail throughout
this paper.

Dust usually carries an electric charge, so, it seems
likely that there might be regions of increased dust particles
more or less tied to magnetically confined areas.
An analog for such accumulations could be the Coronal
Mass Ejections in the solar wind.
The {\it Cassini} mission offers an excellent opportunity to
search for similar dust clouds, more so, as the existence of the
plasma blobs has already been verified.
The source of plasma enhancements is still unresolved, but a
plenty of mechanisms were debated.
We will briefly review five of them in the next section.
Thereafter, we show an example for a ``dust cloud'' and a
``plasmoid'' as well as a connection of both.
For the coupling of these two types, we empirically present
a multi-instrument evidence.


\section{Plasma and dust models for the magnetosphere of Saturn}


\subsection{Sputtering}

Among the first accounts on plasma concentrations at Saturn,
\cite{frank-etal_1980} came forward with ion observations
made by {\it Pioneer 11} in 1979.
They identified a torus of cold hydrogen (H$^{+}$) and oxygen
ions (O$^{2+}$ and O$^{3+}$) in the distance range of
$\approx$4--16 $R_{\rm S}$.
They proposed a sputtering mechanism from two sources:
the Saturnian rings as the primary source as well as the ice
moons Dione and Tethys as secondary sources.
Almost three decades later, the Cassini Plasma Spectrometer (CAPS)
confirmed the existence of the plasma injections from these
moons (\cite{burch-etal_2007}).
The injections are very localized, while the inherent
electrons have pitch-angle distributions characteristic for
a trapped population.

Meanwhile, the geologically active moon Enceladus became known
as a significant contributor of dust, water ice, salty
contaminations, and charged particles, all of which would
further disintegrate in the tenuous environment
(\cite{porco-etal_2006}).
Enceladus attracted more notice and is considered as the main
source of ejecta now.


\subsection{Titan wake} \label{ch:titanwakes}

The second explanation on those density enhancements was
associated with a plasma wake from Titan.
Although lacking a magnetic field of its own, Titan's atmosphere
is known as an ion source (\cite{bridge-etal_1981}).
From {\it Voyager}'s plasma measurements, \cite{hartle-etal_1982}
exemplified that the wake will be produced as a result of the
interaction of Saturn's magnetosphere with Titan's
induced ionosphere,
in a similar way as the interaction of Venus with the solar wind.
The wake would be draped in the moon's quasi-magnetospheric tail,
and it coincides with an abrupt lack of high energy electrons
($>$700 eV), called the electron ``bite-out'' region.
As evolving on its backside sector, the wake arises preferably
on the sunward side when Titan is exposed stronger to the
variable shock front of Saturn's magnetosphere
(Fig.~6 of \cite{hartle-etal_1982}).
In some rare cases, Titan would entirely leave Saturn's
asymmetric magnetic field and be shocked directly by the
solar wind.
Then, the solar wind pressure is likely to evoke reconnection
processes.
Unlike Venus or Mars, all these interactions turn out very
complex in the special case of Titan, for its atmosphere is
affected by both Sun and Saturn (see \cite{bertucci-etal_2011}).

A plasma wake can acquire a meandering shape due to its
stochastic radial motion (expansion, contraction of the entire
magnetosphere of Saturn) in response to the changing solar
wind pressure.
If it remains identifiable for a time longer than a planetary
rotation, it may wrap up itself and stay in the vicinity of Titan.
Thus, the {\it Voyager} spacecraft could have intercepted the
same plasma wake more than once and mocked several density
enhancements (\cite{bridge-etal_1981}).
Two decades later, the {\it Cassini} spacecraft confirmed
different plasma populations during two close Titan flybys
in 2004 (\cite{szego-etal_2005}).
The populations were found drifting as far as 1 $R_{\rm S}$
(or 20 Titan radii) from the moon itself.
However, it was not discernible whether the same furled
structure was traversed, or different particle flows were
generated at different times, or even both.

%
\subsection{Detachments from the magnetospheric sheet}

A third model was presented by \cite{goertz_1983}.
He pointed out that the observed plasma ``islands'' could be
detached from the magnetospheric sheet of Saturn.
This idea is motivated by \cite{hill_1976}
on flux-tubes in a centrifugally distorted magnetic field.
A flux-tube becomes unstable when its plasma content is a
decreasing function of the distance from the spin axis.
%
The theory was successfully applied to the outer magnetosphere
of Jupiter.
For Saturn, such extensions would happen on the boundary of the
magnetospheric sheet to the outer magnetosphere, i.e.\
at about 16 or 17 $R_{\rm S}$.
\cite{goertz_1983} compared the enclosed bubbles of
tailward-moving plasma, which he called ``plasmoids'', with the
separating magnetic blobs already known from the Earth's
magnetosphere.

In the same year, \cite{sittler-etal_1983} presented a
comprehensive analysis of the plasma observations and sketched
a magnetospheric environment around Saturn with three
fundamentally different regions:
the inner plasma torus ($<7\; R_{\rm S}$),
the extended plasma sheet (7--15 $R_{\rm S}$),
and a hot outer magnetosphere ($>15\; R_{\rm S}$)
which extends right down to the magnetopause.
The {\it Voyager} data implied that the disturbances display
an anti-correlated nature of electron properties
(density and temperature):
Hot and more tenuous electrons drift from the outer magnetosphere
inwards, while a cool and rather dense electron component
from the inner torus migrates outwards, combined with heating.
So, there exists an inward and outward transport of plasma
electrons.

The morphology of the magnetospheric profile was essentially
confirmed by {\it Cassini} (\cite{arridge-etal_2007}).
The magnetometer observations also revealed those plasmoids.
The examples presented originate from a radial distance of
$\approx$ 22 $R_{\rm S}$. 
The structures were interpreted as multiple crossings of a rippled
or displaced current sheet caused by solar wind variations.
%
Three more plasmoids were identified in the magnetic tail
at distances as far as 40 $R_{\rm S}$ (\cite{hill-etal_2008}).

The model of plasma detachments does not necessarily contradict
the Titan-associated wake model above (Sect.~\ref{ch:titanwakes}),
for both are consistent with gradual aging effects.
\cite{sittler-etal_1983} estimated the time scale for the
dispersal to be a few Kronian rotation periods.
As \cite{goertz_1983} pointed out, a Titan plume would remain
stable as long as it moved very slowly.
When speeding up, the velocity shear across the wake can generate
a Kelvin-Helmholtz instability which will lead to a mixing of
the slow wake and fast background plasma.
Also, the centrifugal instability will cause a rapid
dispersal in radial direction.
It is even likely that the combined effects account for the
diffusion.  

The sources of these plasma injections remain unknown, but
they seem to be randomly distributed in both local time and
Saturnian longitude.
Another argument in favor are the matching values of the
peak densities of both plasma blobs and in the plasma sheet.

Last but not least, \cite{sittler-etal_1983} showed that the
attenuation of electron energies was correlated with
micron-sized dust particles.
In particular, the authors referred to the inner plasma torus
inside of 5 $R_{\rm S}$, 
though such signatures would easily apply to any diamagnetic
shielding of neutral material in general.
The range of the attenuation will be approximately equal to
the diameter of the dust particles.


\subsection{Hyperion as dust source}

A new angle of view was thrown in by
\cite{bana-krivov_1997}.
Using numerical simulations, the authors argued that Hyperion,
at 24 $R_{\rm S}$, would possibly serve as a dust supplier
for Titan.
Initially, the dust particles of a few micro-meter in size,
originating from Hyperion,
are locked in a 4:3 mean motion resonance with Titan and forming
a stable dust belt.
That resonance will be destroyed by the solar radiation pressure,
and, to a much smaller share (less than 1\%),
by the plasma drag force (\cite{krivov-bana_2001}).
Once the resonance is broken, the orbits of the dust
particles become unstable and stratify themselves by mass:
the larger ones of $\approx 5 \mu$m segregate inwards
and collide with Titan,
the smaller ones will escape out of the system.

Since both Hyperion and Titan move close to the boundary of
the day-ward magnetosphere of Saturn,
the region between the two moons is filled with low density,
hot, and subsonic plasma.
In this region, the interaction occurs between the dust particles
and the plasma:
The much faster moving plasma of ions exposes a drag force
on the dust grains (\cite{bana-krivov_1997}).
%
However, at the time of designing the theory, {\it Cassini}
was still en route to Saturn and the parameters of the dust
as well as the assumed yields of ejecta from Hyperion were
poorly known.
Many assessments were uncertain.

Later, \cite{kennedy-etal_2011} looked for such dust swarms in
the far-off field at Saturn ($>$100 Saturnian radii, $R_S$)
deploying observations of the {\it Spitzer Space Telescope}
in the infrared.
A large-scale cloud, that could be attributed to an irregular
satellite or other cosmic origin, was not found definitely.


\subsection{Two-cell plasma convection}

A fifth approach to explain the density enhancements in the
outer magnetosphere was presented by \cite{gurnett-etal_2007}
when studying the Saturn kilometric radiation (SKR).
The SKR is an intense radio emission from the auroral zones,
analogous to the auroral radiation on Earth. 
Their model goes without artificial relations to moons
or their ambiences.
The heart of  \cite{gurnett-etal_2007}'s concept was a
two-cell convection emerging
between the neutral gas torus at the orbit of Enceladus and
the magnetospheric plasma sheet.

The mass loss from the geysers of Enceladus (distance:
3.95 $R_{\rm S}$) feeds the neutral gas torus all the way
along its orbit.
As the plasma passes outward through the inner edge of the
neutral gas torus, it picks up newly ionized particles from
the torus and thereby increases its density.
The density remains lower on the opposite longitude, thus,
the centrifugal forces become different.
This difference drives a large convection cell on either
flanks of the planet \cite[see Fig.~3 of][]{gurnett-etal_2007}.
The rapid rotation of the plasma disk and the constrained
motion of the particles to the magnetic field lines act
additionally to concentrate the plasma near the equatorial
plane. 

As both convection cells rotate, their outflow produces
perturbations that drift into the outer magnetosphere.
The associated perturbations in the magnetic field develop
a phase lag.
By the time these perturbations reach the magnetopause at
$\approx$20 $R_{\rm S}$, they appear at the morning side
of the planet, where the SKR would be generated.
This would be consistent with the fact that the electron
density is often different on the inbound and outbound
arcs of the {\it Cassini} trajectory, in particular in
the vicinity of Enceladus (\cite{gurnett-etal_2007}).
Thus, the period of the SKR modulation proved to be locked
to the variation of the plasma density in that inner
region of the magnetosphere.

In fact, {\it Ulysses} was able to sense the SKR from Saturn.
The observations by \cite{galopeau-lecacheux_2000} showed
that the SKR period lengthened by
1\% on an annual time scale during the years from 1994 to
1997.
For this long-term variations, \cite{gurnett-etal_2007}
suggested the seasonal effect of the altering solar
inclination angle which affects the conductivity of the
plasma disk:
The illumination of the Southern hemisphere increases
the conductivity of its ionosphere,
which, in turn, governs a number of other phenomena
including the poleward motion of flux, the corresponding
aurora with its SKR, and even connects farther out to the
plasma disk.

Almost the same conclusion was given by
\cite{goldreich-farmer_2007} who found that the SKR must
be supported by currents external to the planetary body.
They also rested the source of the varying SKR upon the
coupling of the outflowing plasma and magnetic field
close to the orbit of Enceladus.
A current would produce a non-axisymmetric component to
the intrinsic magnetic field of the planet, which itself
should be symmetric because of its perfect axis alignment
with the rotational axis.
%
To sum up this model:
The external magnetic field suffers a break in symmetry
due to the centrifugally driven convection.


\section{Data basis and orbits}

The {\it Cassini} orbiter is in the favorable situation
to exploit the dynamics in the magnetosphere in detail.
The multi-instrumental view opens up a new picture to the
precise topographical survey of the magnetospheric regions.
We adopt the model of radial distances by \cite{sittler-etal_1983}
which was improved by \cite{andre-etal_2008}.

The key parameters of our study are provided by the MAPSview
database.
We employed the following instrumental data:
\begin{itemize}
\item CDA: impact rate $r^{\prime}_{\rm all}$ of the registered
   dust events per 64 s, see \cite{khalisi-etal_2015} for details.
\item CAPS: electron density $n_{\rm e}$,
   electron velocity $v_{\rm e}$,
   and pressure $P_{\rm e}$ (EMNT parameter);
   and the same for ions, $n_{\rm i}$, $v_{\rm i}$, $P_{\rm i}$
   (IMNT parameter), if present.
\item MAG: strength of the magnetic field $\left|{\bf B}\right|$
  plus its three components $x$, $y$, and $z$ in the kronocentric
  solar-magnetospheric (KSM) coordinate system.
\item RPWS: qualitative radio signals in the frequency bands of
  1 Hz, 10 Hz, 100 Hz, 1 kHz, and 10 kHz.
\item TRAJ: additionally, the current position of {\it Cassini}
  from Saturn.
\end{itemize}
Most parameters have a resolution of $\approx1$ min of time,
except some very few cases when the instrument was out of its
nominal operation.
The CDA and CAPS are directional instruments,
and their data has to be corrected for the current pointing.
RPWS and MAG are not reliant on the spacecraft attitude and
have the advantage of a continuous measurement of their
respective signals throughout the orbit.
In particular, the components of the {\bf B}-vector give
important clues to the alignment of the magnetic field.


%
\begin{figure}[t]
\centerline{\includegraphics[width=\columnwidth]{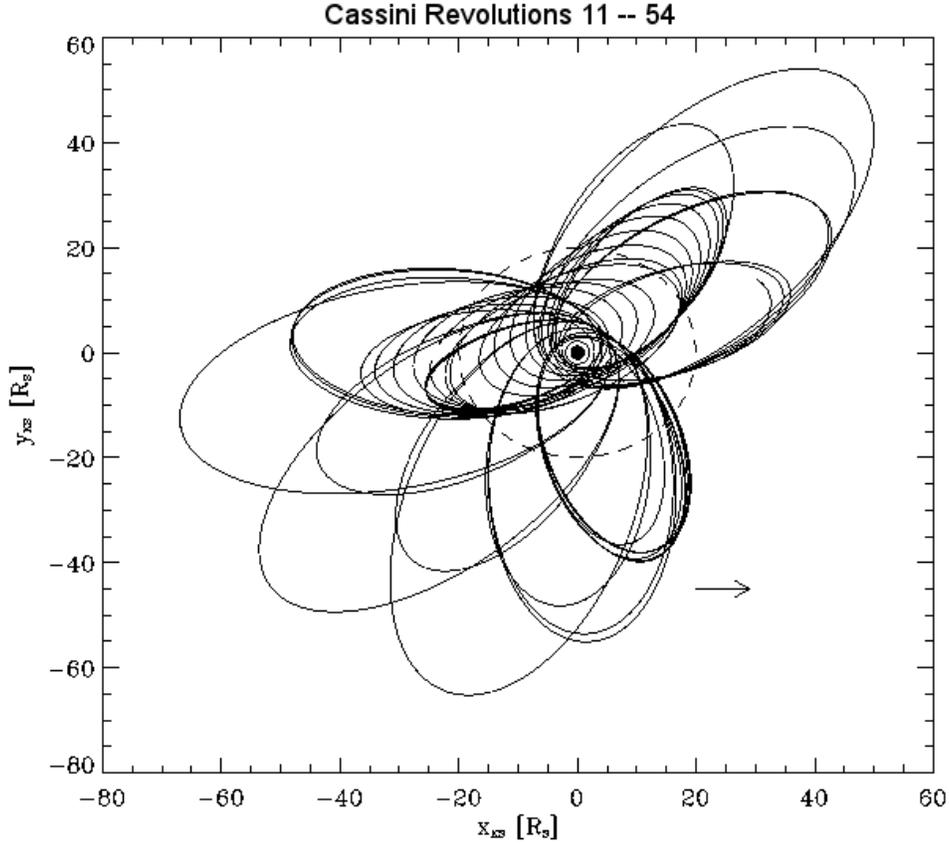}}
\caption{Projection of {\it Cassini}'s Revolutions
\#011--054 (September 2005 to December 2007) on the
ring plane (top view).
The arrow points in the direction of the Sun.
The dashed circle marks the orbit of Titan.}
\label{fig:allrevolutions}
\end{figure}
Our analysis spans the {\it Cassini} Revolutions \#11--54
(DOYs 186/2005 till 361/2007), a total of 905 days.
The state of the Saturnian magnetosphere was evaluated by
\cite{achilleos-etal_2008} for the first Revolutions \#7--14,
which had an inclination of $i = 21^{\circ}$.
We supplement their analysis by 40 more revolutions and put
our focus to the equatorial plane ($i < \pm5^{\circ}$).
Figure~\ref{fig:allrevolutions} shows the projection of the
orbits onto the ring plane in a Saturn-centred coordinate
system.
The Sun was to right hand side (arrow), illuminating the Southern
hemisphere;
it was below the plane at an angle of $\approx$20--10$^{\circ}$
getting increasingly shallower.

The orbital inclinations of {\it Cassini}'s trajectory changed
during its tour.
The equatorial plane was traversed at the dusk side
for the Revolutions \#15--26,
while the line of apsides turned slowly to the magnetotail.
At the interim Revolutions \#27--46 the orbits suffered from
high inclinations ($i > 25^{\circ}$), which we redlined from
this particular analysis.
When the orbits were re-tilted to the ring plane for the
Revolutions \#47--52, {\it Cassini} already transited the
evening side of the magnetosphere.
A disturbance from Jupiter's magnetic field was not given,
for it had outpaced Saturn back in 2000 and, now, was a quarter
of its orbit ahead.
Also, the solar activity passed its maximum in 2000, and
was subsiding.
Therefore, such ``external effects'' can be neglected with
regard to the years 2005 to 2007,
and even so at the distance of Saturn.


\section{Topography of the magnetosphere}

Our approach was such that we firstly determined the state
of the magnetosphere at each revolution.
The electron density in the solar wind reveals stable
at values of $n_e \approx 10^{-1}$ cm$^{-3}$,
as are the electron velocities $v_e$ at $\approx$20--40 km/s.
Though the electrons are thermal velocities, they provide
continuous data and a resolution on 1-min-scale, while
the ion data are often missing in the data base.
The electron data proved benefical to distinguish the borders
of the magnetospheric regions.
The interplanetary magnetic field strength is usually
$\left|{\bf B}\right| \approx$0.5--1.0 nT at the distance
of Saturn.

The magnetopause of Saturn is identified best by a sharp drop
of $n_e$ and a steep rise of $v_e$.
Simultaneously, $\left|{\bf B}\right|$ jumps to 4--5 nT
and the $B_z$-component turns negative.
The B-values at the entry of the magnetopause do vary strongly
upon local conditions.
In a number of cases, when the spacecraft sojourned near the
magnetopause, we observed a swaying magnetic boundary
(multiple crossings of the magnetopause);
or an accumulation of dust particles at the bow shock;
or various detached plasmoidal regions;
or a sudden depletion of electrons.
The whole magnetopause appears very vivid, complex, and looks
different at almost each passage.
On the other hand, some dynamical phenomena show a surprisingly stable
pattern that remain as long as several revolutions of {\it Cassini}.
Therefore, the identification of some features inside the
magnetosphere turns out ambiguous.
\cite{andre-etal_2008} confined four regions which we re-examined
in our broader sample:
\begin{itemize}
\item the innermost plasma disk, 
\item outer plasma disk, 
\item plasma sheet (at radii $\approx$7--16 $R_{\rm S}$),
\item and the hot outer magnetosphere reaching out to the magnetopause.
\end{itemize}
The latter region has been tagged as ``magnetic cushion''
recently (\cite{delamere-etal_2015}),
as the processes between the current sheet and the
magnetopause are not well understood in case of Saturn.
The cushion behaves like a reservoir of magnetic flux balancing
the centrifugal and mechanical stress when the magnetodisc
re-structures itself dynamically under the variable solar
wind conditions.

When the spacecraft entered any of these magnetospheric regions,
the magnetometer (MAG) and plasma instrument (CAPS) displayed
some characteristics that are summarized in
Table~\ref{tab:magsignatures}.
The features represent a very rough indication and
are based upon the comparison of many orbits.

\noindent
\begin{table}[t]
\caption{Magnetospheric features and their characteristics for
   distinguishing the regions amid the instrumental data.} \label{tab:magsignatures}

\medskip
\begin{tabular}{|l|p{5cm}|p{5cm}|}
\hline
             & MAG & CAPS \\
\hline \hline
\rowcolor{grey10}
solar wind   & $\left|{\bf B}\right| \approx$ 0.5 -- 1.5 nT
             & $n_e \approx 10^{-1}$ cm$^{-3}$, stable \\

\rowcolor{grey10}
             &                                      
             & $v_e \approx 20-40$ km/s \\ 

\hline \hline
magnetopause & $\left|{\bf B}\right|$ abrupt rise to $\approx$4--5 nT
                   & $n_{\rm e}$ abrupt drop by 10--100$\times$\\
             & $B_z$ sharp kink to negative
                   & $v_{\rm e}$ upward jump by 10$\times$\\
\hline
magnetic cushion& co-rotating $B_y$
                   & $n_{\rm e} \ll 10^{-1}$ cm$^{-3}$, fluctuating\\
             &                                 
                   & $v_{\rm e} \gg 10^2$ km/s, unstable\\
\hline
plasma sheet & all $B$-components $\approx$ const.
                   & $n_{\rm e}$ small rise, less fluctuating \\
             & low variance of $\left|{\bf B}\right|$, stable
                   & $v_{\rm e}$ small drop\\
\hline
outer plasma disk & ($B_y$ first minimum)
                   & $n_{\rm e}$ exceeding 1 cm$^{-3}$\\
             & 
                   & $P_{\rm e} \approx 0.5$ eV/cm$^{-3}$ \\
\hline
inner plasma disk & ($B_y$ second minimum)
                  & $n_{\rm i} > 100$ cm$^{-3}$\\
             & 
                  & $P_{\rm i}$ rising by factor 5--10\\
\hline
\end{tabular}
\end{table}

\begin{figure}[t!]
\centerline{\includegraphics[width=\columnwidth]{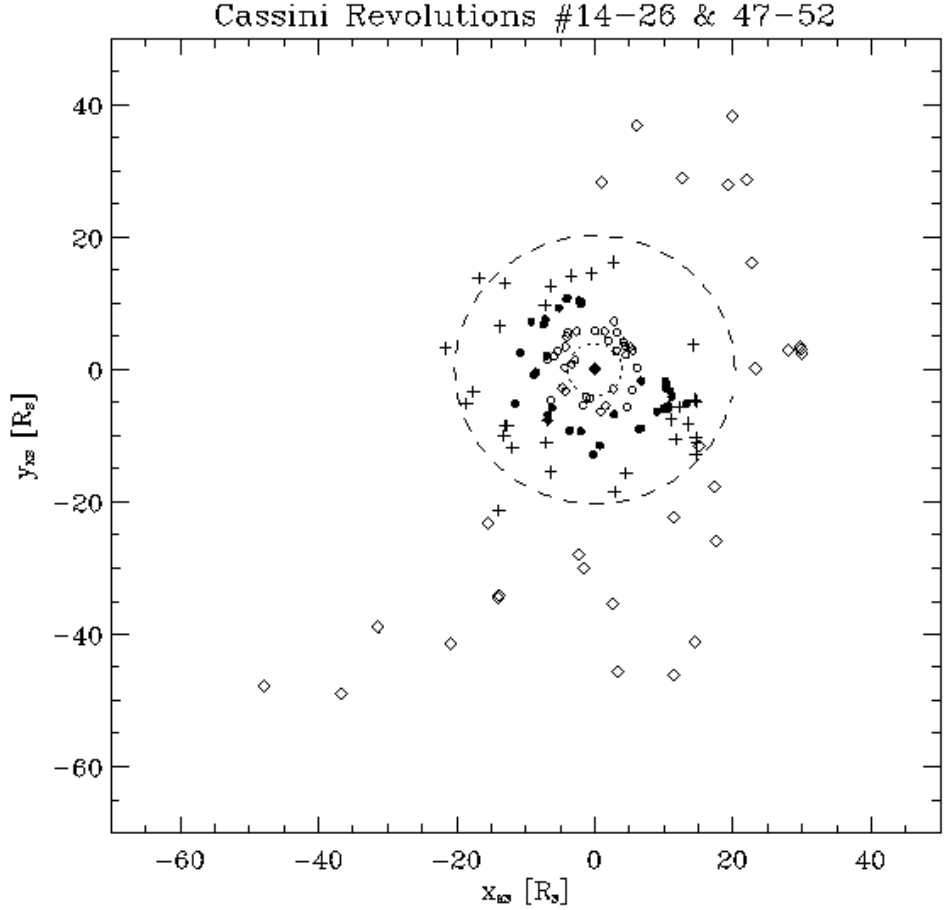}}
\caption{Crossings of the magnetospheric features with the
trajectory lines of Figure~\ref{fig:allrevolutions}
omitted.
Shown are the boundaries of the magnetopause (diamonds),
plasma sheet (black circles),
outer disk (crosses),
and the inner disk (small points),
with all revolutions performed close to the equatorial plane.}
\label{fig:equatorial-magnetopause}
\end{figure}

\begin{figure}[t!]
\centerline{\includegraphics[width=\columnwidth]{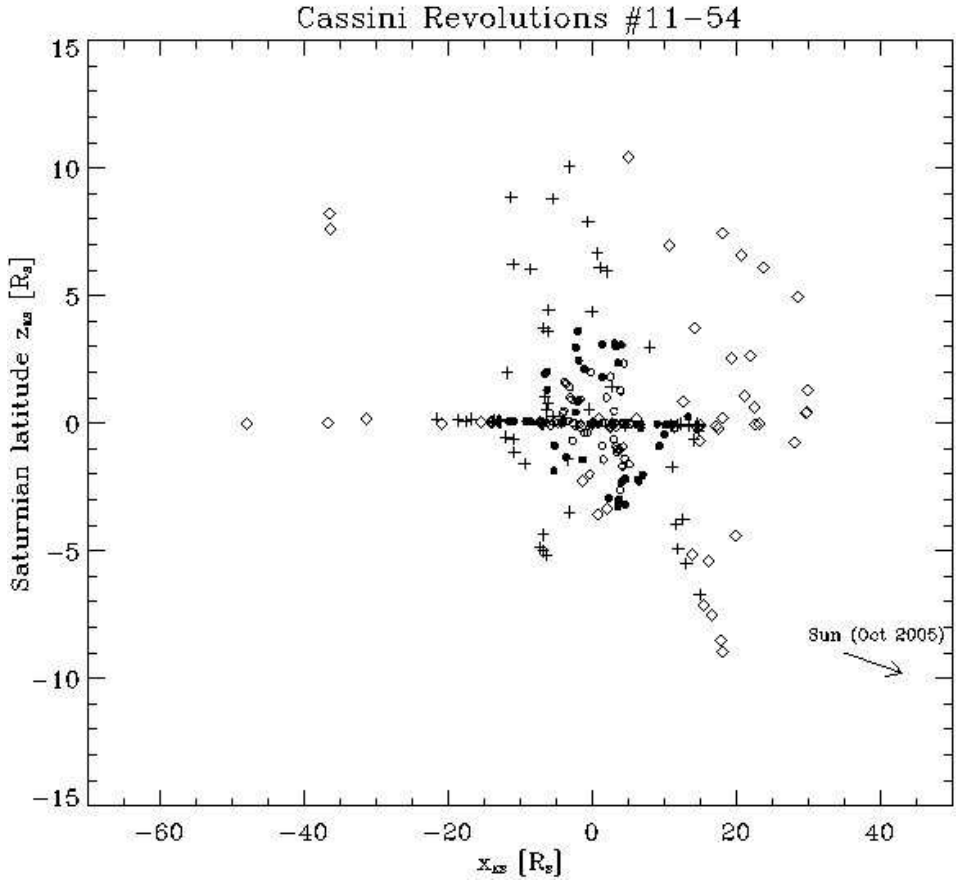}}
\caption{Vertical profile of the magnetopause (diamonds),
the plasma sheet (black circles),
the outer disk (crosses),
and the inner disk (dots)
for all Revolutions \#11--54 (July 2005 -- Dec 2007).}
\label{fig:magheight}
\end{figure}

Figure~\ref{fig:equatorial-magnetopause} displays the points
of entering and leaving the respective region during the
Revolutions \#15--26 and \#47--53, as {\it Cassini} flew its
first two bunches of equatorial orbits (inclination $i < 5^{\circ}$).
In an edge-on view, Figure~\ref{fig:magheight} shows these
features for all Revolutions (\#11--54).
The Sun reduced its angle of illumination from $-20^{\circ}$ (arrow)
to $-10^{\circ}$ during the time span considered in this study
(mid-2005 to 2007).

\section{Examples of dust clouds}  \label{ch:dustclouds}

We searched the whole data sample for patterns that repeatedly
appeared in three or more subsequent revolutions.
Particularly, we put our focus on dust particles of the CDA.
An intriguing pattern was presented by \cite{khalisi-etal_2015}
for the Revolutions 26--29.
Here, we revisit that data and supplement the evidence
in the scope of additional instruments.

%
\begin{figure}[th]
\centerline{\includegraphics[width=\columnwidth]{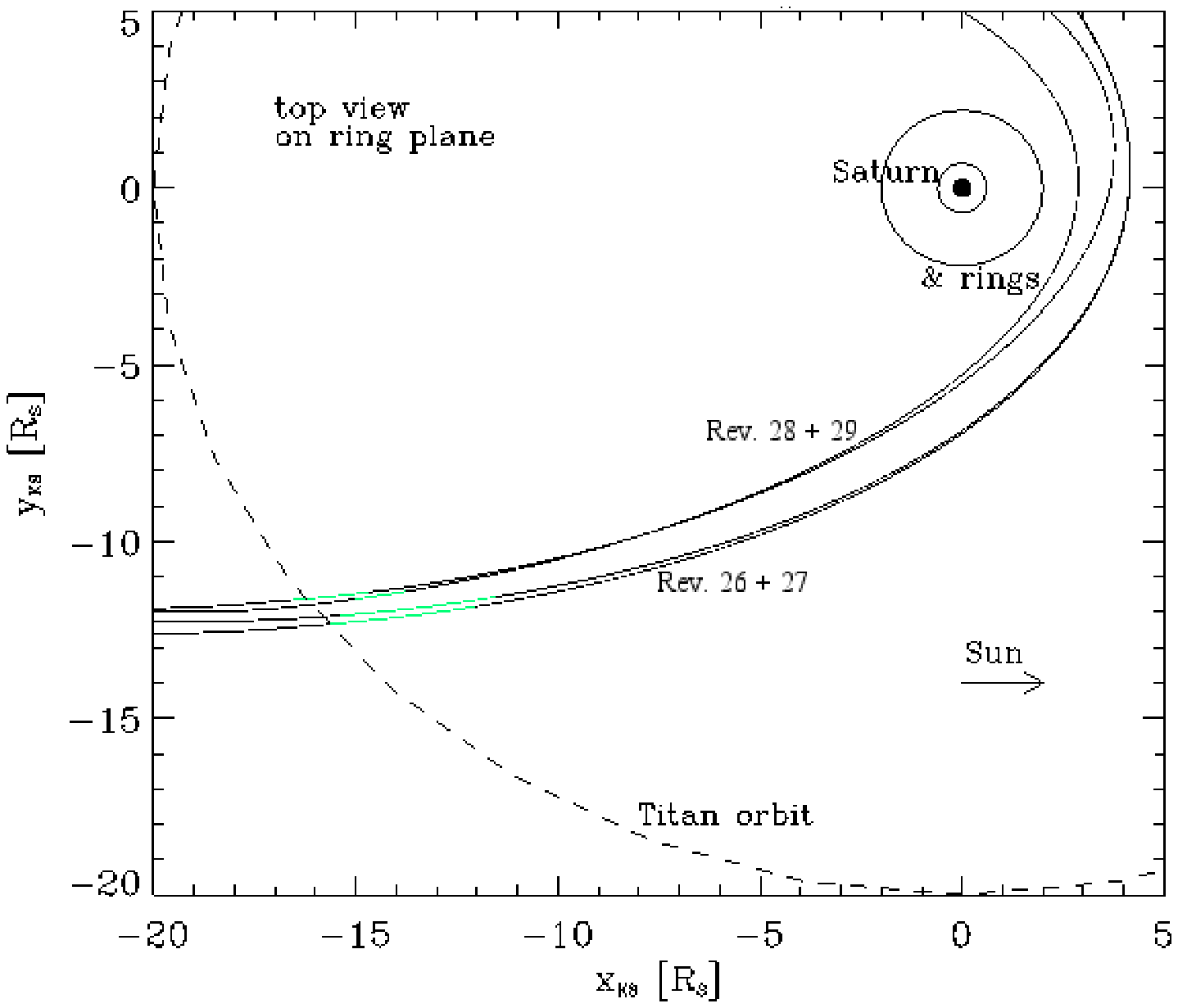}}
\caption{Trajectories of the {\it Cassini} Revolutions
\#26--29 (July--Sept 2006).
The region of the dustcloud is highlighted green.}
\label{fig:orbit-arcs}
\end{figure}

\subsection{Three-peaked dust pattern at Rev.\ 26--29}

\begin{figure}[t]
\centerline{\includegraphics[width=\columnwidth]{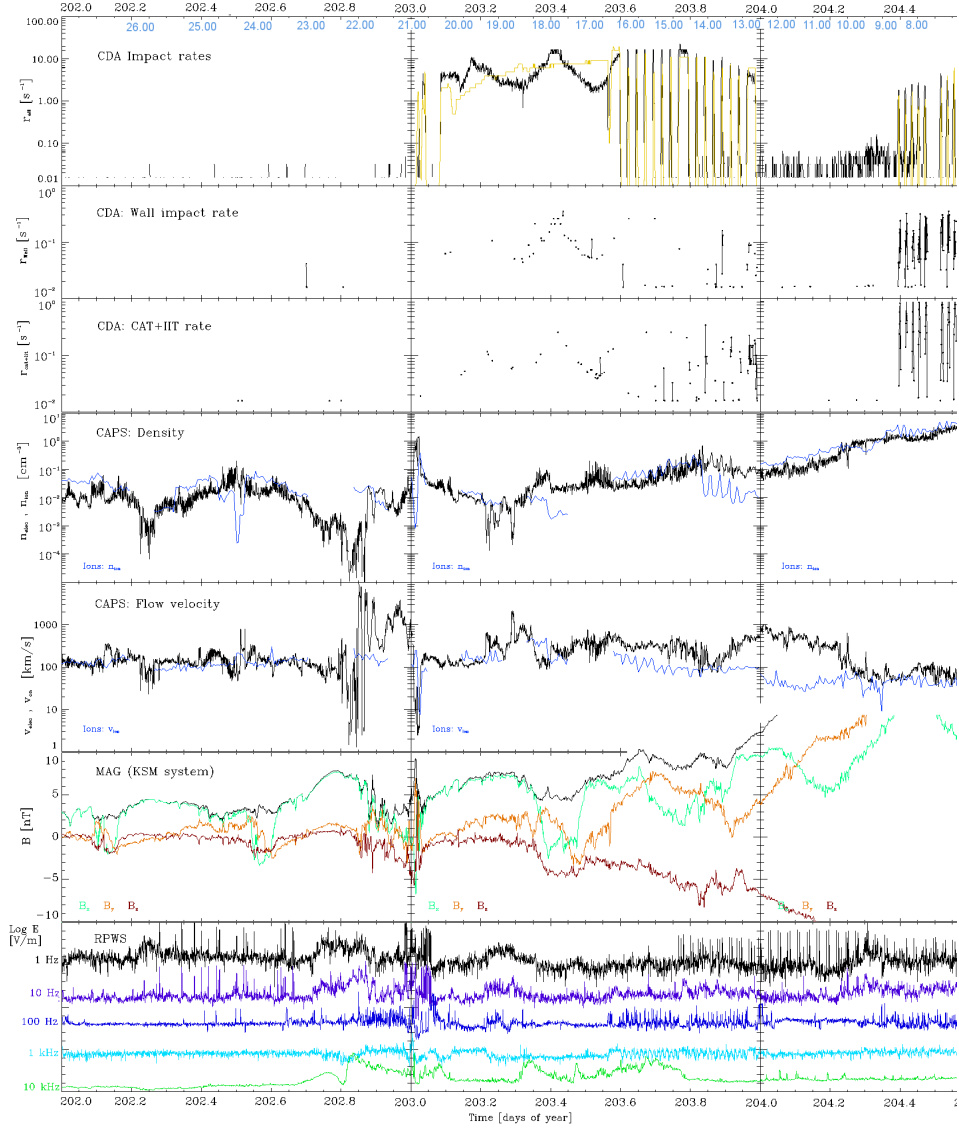}}
\caption{Comparison of four {\it Cassini} instruments for
Revolution \#26 in July 2006.
{\it Uppermost panel:} Total impact rate $r_{\rm all}$
of all impacts on the CDA (black line) as well as the sensitive
area (yellow line) exhibited to the Kepler-RAM.
Distances from Saturn in $R_{\rm S}$ are ticked in
blue colour.
{\it Second + third panel:} Impact rates at the
instrument housing (Wall) and the main targets (CAT + IIT),
respectively.
{\it Forth and fifth panel:} Densities of electrons
(black) as well as ions (blue) and their velocities
from the CAPS instrument.
{\it Sixth panel:} Magnetic field strength
$\left|{\bf B}\right|$ (black) with their spacial
components.
{\it Bottom panel:} Five signals of the lower frequency
bands from the radio and plasma data (RPWS) in arbitrary units.}
\label{fig:cloud-rev026}
%
\end{figure}

{\it Cassini} approached its perikronium in Revolution \#26
down from the magnetotail, while the magnetic field was rather
quiet ($\left|{\bf B}\right| \approx$5 nT),
but slowly rising when approaching the inner regions.
In the far-off space beyond 20 $R_{\rm S}$,
\cite{arridge-etal_2009} identified several crossing events
of the current sheet.
On DOY 203.02 of 2006 a very close flyby of Titan
(T16, distance: 950 km) took place.
This passage was accompanied by a vigorous disturbance in
all instruments (Figure~\ref{fig:cloud-rev026}).
The magnetic data clearly shows how {\it Cassini} crossed the
moon's induced ionosphere.
Right after the passage, the spacecraft changed its sequence of
equatorial orbits and set in for a number of inclined orbits.
On four consecutive revolutions, \#26--29, it traversed
almost the same spot in the Saturnian space
(Fig.~\ref{fig:orbit-arcs}).
On the first of these (\#26), the ring plane of Saturn was
crossed about two hours after that Titan flyby (DOY=203.11).

%

Then the CDA data revealed three conspicuous peaks of impacting
particles at DOY 203.17, 203.41, and a broad one around 203.70
(Fig.~\ref{fig:cloud-rev026}).
The latter one happened during the communication period
with Earth, when the probe carried out ``rolls'' while
scanning the sky for dust impacts.
From the second and third panel of Figure~\ref{fig:cloud-rev026}
it is seen that twice as much impacts were registered on the
instrument housing (so-called ``Wall events'')
than on the main sensitive area (``CAT + IIT events'').
That means that most particles entered the device
from a direction deviating from the Kepler-RAM.
The ratio of both rates would reflect the angle $\alpha$
between the two flows:
%
\begin{eqnarray}
\frac{r_{\rm CAT} + r_{\rm IIT}}{r_{\rm Wall}}
= \cos\alpha .
\end{eqnarray}

A rough estimate yields that the particle cloud
drifts toward $\alpha \approx 25^{\circ}$ relative
to the Kepler-RAM.
Here, we accumulated the few available data

\begin{figure}[t!]
\centerline{\includegraphics[width=\columnwidth]{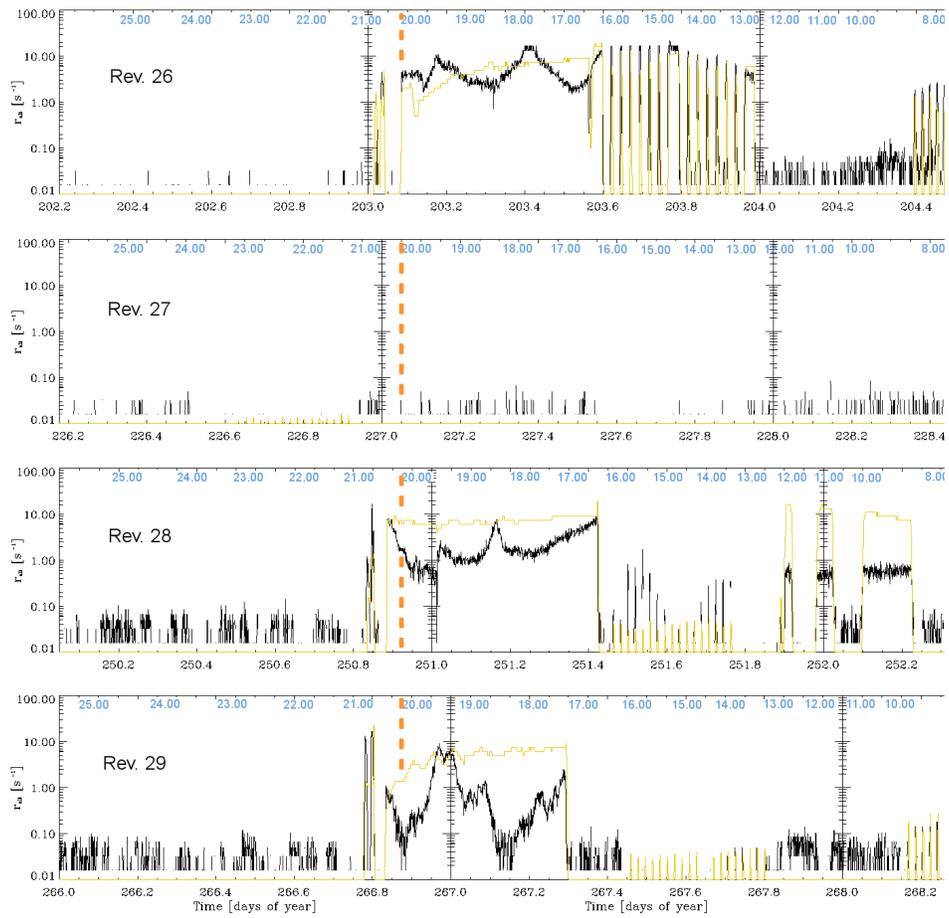}}
\caption{Dust rates and the exposed area (yellow) of the
CDA for the Revolutions 026--029 (July -- Sept 2006).
The panels were aligned to the Titan orbit
(orange dotted line).
The current distance from Saturn is given at the upper
axis with blue marks.}
\label{fig:cloud-cda}
\vspace{-3cm}
\end{figure}
\clearpage

\noindent
points between $T=$ 203.30 and 203.55 to intervals of
30 min and computed the averages.
The second dust peak coincides with a perturbation of
the magnetic field, as can be seen in the sixth panel
of Figure~\ref{fig:cloud-rev026}.
It is also conceivable that a magnetic compression region,
filled with dust particles, could have passed the
spacecraft.

A comparison of the dust rates for the subsequent {\it Cassini}
revolutions is shown in the next Figures.
The second panel of Fig.~\ref{fig:cloud-cda} reveals that
the CDA pointed to an unsuitable direction, away from the
Kepler-RAM.
A few ``accidental'' impacts still made it for measurement
in the time period under consideration, i.e.\
from $T \approx$ 227.10 to 227.55.
An enhancement of dust particles can be sensed.

At the third return to that same spot of space (third
panel of Fig.~\ref{fig:cloud-cda}),
the signals resembled those of Revolution \#26 again.
The three-peaked pattern can be identified at
DOY = 251.02, 251.16, and 251.65.
Finally, at Revolution \#29, the orbit inclination of the
spacecraft had already been increased by some $10^{\circ}$,
thus, {\it Cassini} must have transited the Southern parts
of that dust cloud.

Examining the two front peaks separately, one finds
that both have experienced an apparent compression and
moved outwards (Tab.~\ref{tab:timestamps}).
The prior peak was squeezed by factor of 3, and the
later by 1.5.

\begin{table}[t]
\caption{Time markers T and distances d [in $R_{\rm S}$] for the
two main dust peaks during the orbits 26 and 28, respectively.
See also Figure~\ref{fig:orbit-arcs}.}
\label{tab:timestamps}
\medskip
\begin{tabular}{lrrrrr}
\hline
      & Minimum& Maximum &Minimum &Maximum &Minimum  \\
\hline
 T$_{26}$= & 203.14 & 203.175 & 203.31 & 203.41 & 203.525 \\ 
 d$_{26}$= & 19.90  & 19.64   & 18.60  & 17.80  &   16.86 \\ 
%
\hline
 T$_{28}$= & 251.01 & 251.02  & 251.08 & 251.16 &  251.24 \\ 
 d$_{28}$= &  19.65 &  19.58  &  19.15 &  18.56 &  17.96 \\ 
\hline
\end{tabular}
\end{table}
The second peak seems to approach faster,
pushing the prior.
The front side of the first pile appears steep in
Revolution 26, while its tail runs out shallower.
But then, in Revolution 28, the second pile compresses
the forerunning and could possibly have merged into it
by the time of Revolution \#29.
Since the spacecraft changed its orbit inclination,
it would have traversed a different part of that
cloud then.
From the time stamps of entering and leaving the cloud,
the extent of both patterns can be estimated to
$\approx$82,500 km or 1.36 $R_{\rm S}$.

\subsection{Pulsating magnetosphere at Rev.\ 15--18}

A magnetic reconnection event in the magnetosphere was
observed during the Revolutions \#15--18 in late 2005.
We claim that the process persisted for several days
and left behind disrupted plasmoids.
The data of three consecutive revolutions resemble a
``respirating magnetopause''.
The relicts of that process stayed as long as 35 days --
much longer than the reconnection events we would expect
from Earth's magnetosphere.

\begin{figure}[h]
\vspace{-2cm}
\centerline{\includegraphics[width=\textheight, angle=90]{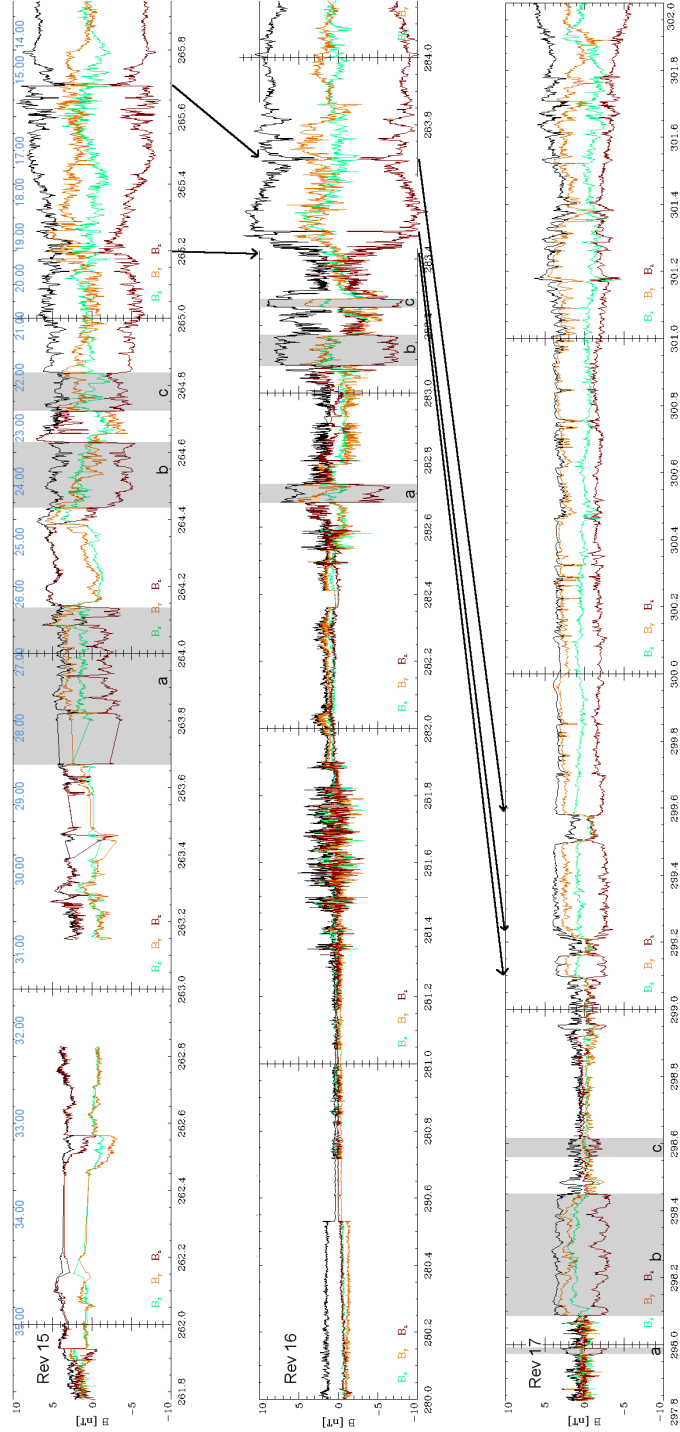}}
\caption{Magnetic field during the Revolutions 15 to 17
at the same spot of Saturnian space.
The distance to Saturn is indicated at the top
in blue.
The modulus of the magnetic field, $\left|{\bf B}\right|$,
is the upper black line.
The shaded areas indicate some similarities of magnetic
data being streched or compressed, respectively.
The arrows show the cut-offs (reconnections) that would lead
to plasmoids in the Revolution \#17, bottom panel.}
\label{fig:plasmoid-rev16}
%
\end{figure}

The three panels of Figure~\ref{fig:plasmoid-rev16} show the
magnetic field data of the Revolutions \#15--17,
and they are adjusted to the distance scale, which is given at
the uppermost axis in blue ticks.
{\it Cassini} was in the equatorial plane with the line of
its apoapsis pointing to the dawn side.
It passed almost the same spot of the Saturnian coordinate space
for the next five revolutions.
The spacecraft had already struck the magnetosphere at distance
$d=$37.77 $R_{\rm S}$ on DOY = 260.98 (not shown here).
The $B_z$-component had switched from positive to negative values
of $\approx \pm5$ nT at various times (shaded gray and marked
with letters a, b, and c),
before remaining negative at distance $d$ = 22.65 $R_{\rm S}$
on DOY = 264.72.
This marked the final entering into the magnetosphere.

At the next return (\#16, middle panel), most of the positive
$B_z$-regions turned to segments of high magnetic turbulence.
They are still visible in the intervals between the gray
areas;
the most conspicuous one appeared still reconnecting during
DOY 281.
The gray areas became detached plasmoids in a compressed
magnetic field with their negative $B_z$-component still
conserved inside the loop.
A new magnetopause started re-forming further inside between
$d$ = 21.50 and 19.50 $R_{\rm S}$ during DOY 283.
The highest spikes of the $B_z$-component evolved to new
cut-offs for two more plasmoids that would be visible at
the next Revolution (indicated by arrows in
Fig.~\ref{fig:plasmoid-rev16}).

At the third return (Rev.\ 17, bottom panel), the interplanetary
magnetic field had created new ambient conditions and made
the magnetosphere cause a vast expansion.
That solitary plasmoid of DOY = 282.70 had shifted by more
than 10 $R_{\rm S}$ outward to DOY = 297.97, and faded.
Other plasmoids were stretched and weakened, while the new
magnetopause established at $d$ = 29.16 $R_{\rm S}$ (DOY = 299.59).

\begin{figure}[th]
\centerline{\includegraphics[width=\columnwidth]{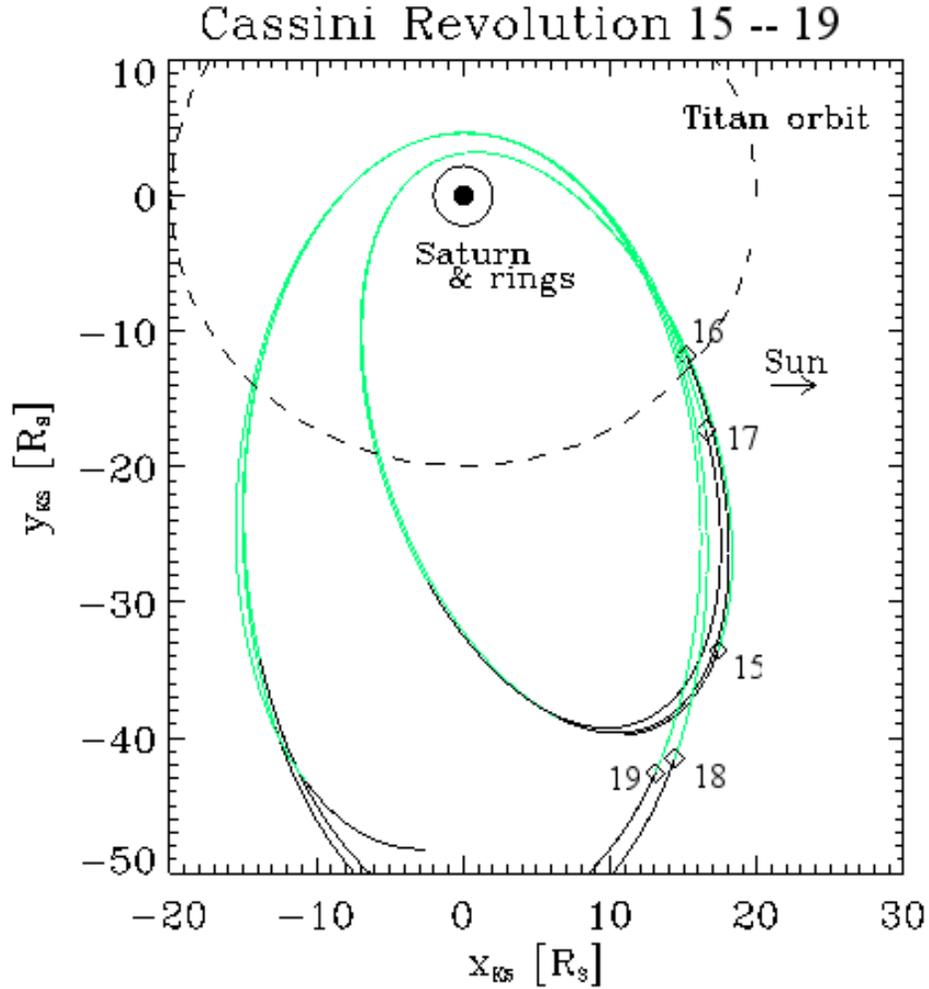}}
\caption{Trajectory for the Revolutions 015--019.
The green arcs mark the space inside the magnetopause.
Diamonds indicate the points of entry into Saturn's magnetosphere.}
\label{fig:orbits015-019}
\end{figure}

The process of restoration of the magnetopause was still
continuing in the subsequent Revolution, \#18.
In Figure~\ref{fig:orbits015-019} we show the pulsation in a
top-view on the equatorial plane.
From the numbers accompanying the entry positions (diamonds)
into the magnetosphere, it is seen how that boundary contracted
and bounced back again at Revolutions \#18 and 19.

The procedure of re-organizing would normally take place within
a few hours or be completed within a day, at most.
$\,$ Therefore it looms unusual 
%
%
\clearpage

\noindent
that such magnetic relicts could
remain for several days, as shown here.
As \cite{fuselier-etal_2014} point out, reconnection events are
identified by two important aspects:
First, their location of occurence (planetomagnetic latitude)
-- this characterizes the type of reconnection in terms of the
shear angle between the magnetosheath and the planet's magnetic
field at the reconnection site.
And second, heated streaming electrons in the magnetosheath near
the magnetopause;
the electron detectors draw an abrupt change of the density
and temperature when passing the magnetopause.
Our explanation of these long-lasting plasmoids
(gray areas in Fig.\ \ref{fig:plasmoid-rev16}) was rather
interpreted as partial crossings of the magnetopause by
the spacecraft.

In general, our scenario disagrees with \cite{bagenal-delamere_2011}
who estimate the outflow of plasma from Saturn's magnetodisc
within one planetary rotation or less than $\approx 1$ day.
Therefore, we cannot exclude being erroneous and dealing with
a different mechanism than suggested here.
However, the similarity of the features in the shaded areas
as well as the turbulence inbetween appear remarkable.
The patterns seem to repeat in the Revolutions \#15--17,
and make us suggest the same magnetic environment.

\subsection{Dust-filled clouds and evacuated holes}

\begin{figure}[th]
\vspace{-2cm}
\centerline{\includegraphics[width=\textwidth,]{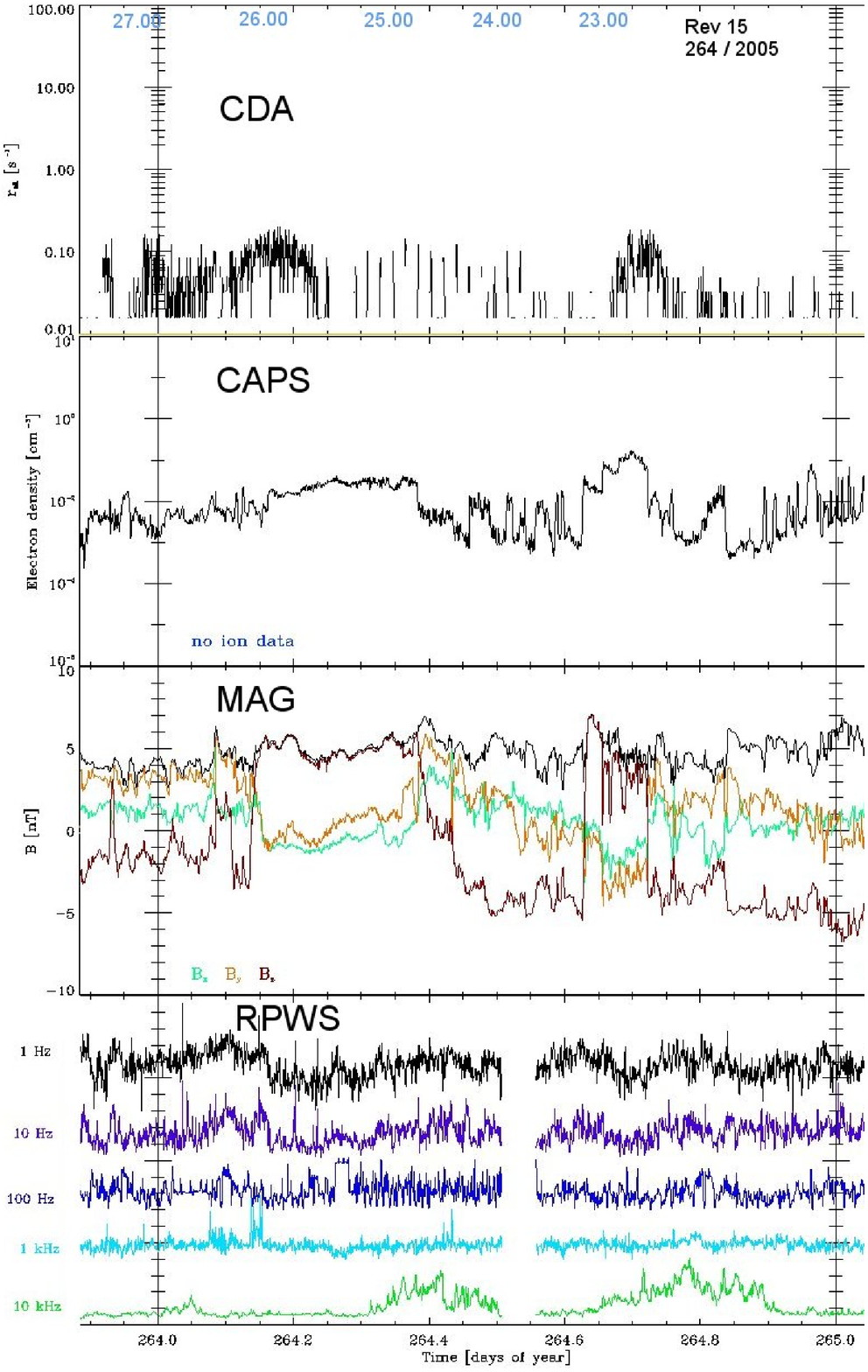}}
\caption{Multi-instrument data before entering the magnetopause
on Revolution \#15 at DOY = 264.72 in 2005.
In the third panel, the $B_z$-component is colored red, while
the modulus of $\left|{\bf B}\right|$ is the upper black line.}
\label{fig:blob-rev015}
\end{figure}

Figure~\ref{fig:blob-rev015} reprises the initial conditions of the
reconnection process in Revolution \#15 in a multi-instrument view.
In the uppermost panel, dust particles gather in those two blobs
at DOY $\approx$264.15--.23 and $\approx$264.67--.75, respectively.
Their impact direction is non-Keplerian, as the pointing of the CDA
(yellow line) is zero.
The electron density (second panel) exhibits high values
matching the temperature of the solar wind;
the magnetic cushion of Saturn is usually colder than these
measurements.
The compression of the magnetic sheath can also lead to higher
thermal electrons, however, also suggesting an incipient
interaction with the solar wind.
These enhancements of dust and electrons coincide with a
northward component of the magnetic field (third panel, red line);
this component will be reconnecting and diminishing thereafter.
The relicts are visible in the next orbit (middle panel of
Figure~\ref{fig:plasmoid-rev16} at DOY=282.55 and 282.76).
The 10-kHz-radio signal of RPWS (green line, bottom panel)
beacons to strong energetic activity.
Both events are an example for a ``dust-filled plasmoid''
indicating a relationship of ionized nano particles with the
magnetic field.

\begin{figure}[t!]
\vspace{-2cm}
\centerline{\includegraphics[width=\columnwidth]{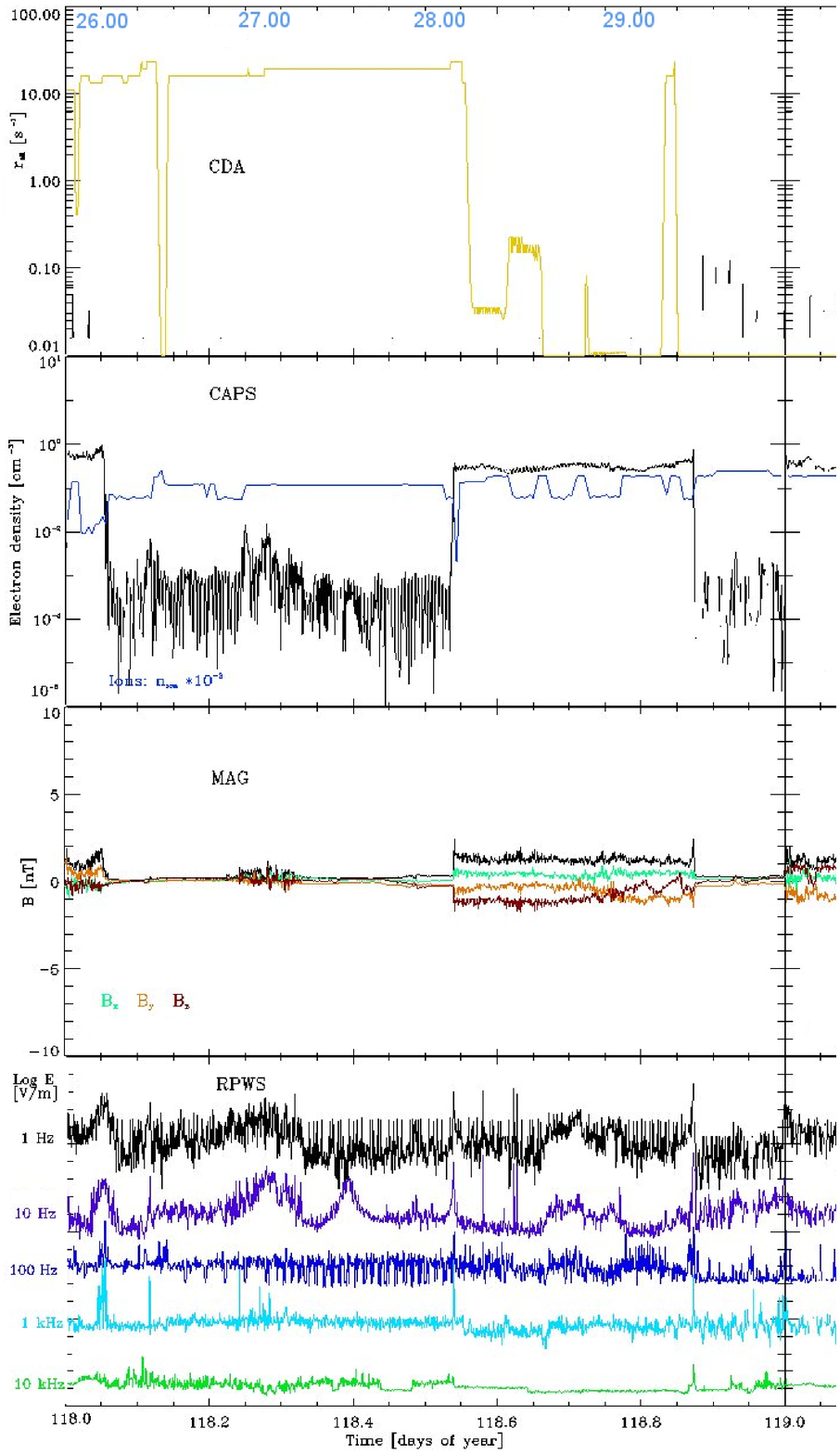}}
\caption{Dust impacts, electron density, magnetic field,
and 5 low-frequency radio signals on Revolution \#43
at DOY 118/2007. Distances from Saturn in $R_{\rm S}$
are indicated at the uppermost axis in blue colour.}
\label{fig:holes-2007118}
%
\end{figure}

The opposite would be some kind of an ``evacuated plasmoid'' and is
presented in Figure~\ref{fig:holes-2007118}.
On DOY 117/2007, {\it Cassini} was on its Revolution \#43, a highly
inclined orbit on the evening side of Saturn.
It had left the magnetosphere on its outbound leg and was
about 10 $R_{\rm S}$ above the ring plane.
In the midst of the solar wind, there appeared patches of
extremely low magnetic field:
The first is seen at DOY = 118.06 with relicts of a magnetic
turbulence centered on time stamp DOY = 118.28,
and the second one starting at 118.88 till the end of the day.
The CDA shows no dust impacts, except a few solitary hits.
However, the pointing was excellent towards the Kepler-RAM and
off, again.
CAPS also suggests two hole-like features with the electron density
(blue line) being depleted by three orders of magnitude.
A comparison of the MAG-data revealed that the $B_z$-component
had a similar northward orientation, as discussed in the section
above, at the respective locations during the previous orbit.


\section{Discussion}

The dust cloud of Revolutions \#26--29 seemed to drift
at a velocity of $\approx$10--14 m/s in space.
Under the assumption that Enceladus is the most significant
source for permanently pouring solid material into the
Saturnian space, such a cloud would need about 2.5 years
to migrate from its origin to the scene of observation
(\cite{khalisi-etal_2015}).
The frequent transits of the icy moons like Tethys, Dione,
and Rhea would quickly have torn apart such a structure.
The particles would be distributed along the way or stick
to the surfaces of these moons.
Various other effects like shock waves, gravitational
drags, evaporation, and Kepler shear will also lead to a
fast disruption of the cloud on the orbital time scale.
Only debris larger than $\approx1$ mm in size may survive,
though such large particles are not recorded.

Along with the example of a dust-filled magnetic plasmoid,
we suggest the following mechanism:
Nano-sized dust particles are ionized at their origin,
which would be the orbit of Enceladus.
This would happen due to UV-radiation or energetic jet
streams.
The particles become trapped inside a ``magnetic cage''.
Then, they will be pushed down the magnetic flow as in the
model of two-cell convection by \cite{gurnett-etal_2007}.

Leaving behind the plasma sheet at $\approx$17 $R_{\rm S}$,
these ``cages'' become isolated and turn into ``plasmoids''
as introduced by \cite{goertz_1983}.
The plasmoids would carry the pack of tiny particles all
the way through the magnetic cushion.
Beyond the magnetopause they are subject to reconnection
effects with the northward-oriented component of the solar
magnetic field.
The plasmoids burst and liberate their interior.
As the swarm disrupts, the nano-particles get integrated into
the ambient medium.
The smallest particles vanish first, while the inert ones
stay for as long as 40 days, as shown in Figure~\ref{fig:cloud-cda}.

\cite{jackman-etal_2014} listed some 100 plasmoids and
other features in the magnetotail during 2006.
Most of them lasted a few minutes, and the longest
$\approx$1.5 h.
The authors consider an underestimation of the plasmoid size
by a factor of 4--8. 
Our particular dust enhancement of Figures \ref{fig:cloud-rev026}
and \ref{fig:cloud-cda} is not connected to any of their
incidents.
The large-scale comparison of orbits leaves us with data
being almost 20 days apart.
This is usually far too long for plasmoids to exist,
therefore, we cannot present unambiguous footage,
but a suggestion to further work.

\clearpage

It may be noteworthy that \cite{krueger-etal_2010} also
reported of surprisingly high impact rates of dust at
Jupiter, when the {\it Galileo} spacecraft was occasionally
at far-off distances from the planet.
Being outside of the Jovian magnetosphere at
$\approx$280 $R_{\rm J}$ or 0.13 Astronomical Units,
an enhanced dust emission was seen and interpreted as signatures
from the innermost Galilean Moon Io or its torus, respectively.
That period coincided with neutral gas production from the Io
torus, such that the dust particles would couple to both
the gas and the magnetic field.
However, the coupling mechanism was not know in detail.

An analogy can be drawn to a Coronal Mass Ejection (CME)
that contains hot ions from the solar atmosphere or a
prominence, in particular.
As \cite{ragot-kahler_2003} pointed out, the electromagnetic
force becomes most effective
for particles smaller than 100 nm in size (their Figure 2).
At the planetary distances analyzed by us, the magnetic field is
much too weak for retaining the larger particles.
But in the immediate vicinity of Saturn, $\le$2 $R_{\rm S}$, the
electrostatic interaction prevails, as the phenomenon of the dark
spokes on the rings of Saturn show (\cite{gruen-etal_1992}).
There, even micro-sized particles are caught by the strong
magnetic field and forced to shift their motion.

In this paper, we suggested a heuristic mechanism that would
carry a swarm of tiny particles over large distances through the
magnetosphere of Saturn.
However, not everybody supports the results presented here.
A detailed mathematical model will be necessary to verify
that view.
The data of the CDA does provide the empirical basis for this.
Even more such cases as in Figure~\ref{fig:cloud-cda} may
be hidden in the legacy of {\it Cassini} and await their
discovery.


\section*{Acknowledgments}

We thank the Klaus Tschira Foundation, Heidelberg, for
the financial support for this study from 2010 to 2013.
The Max-Planck-Institute for Nuclear Physics, Heidelberg,
enabled to finish the work.
Thanks also to Jeffrey Kopmanis, Uni Michigan, for the
help with the MAPSview data base;
Harald Kr\"uger and Norbert Krupp for discussions;
and the referees for their critical comments that led to
significant changes.


%
%
%
%
%


\section*{References}


\begin{thebibliography}{99}


\bibitem[Achilleos {\it et al.}(2008)]{achilleos-etal_2008}
Achilleos N., Arridge C.S., Bertucci C., Jackman C.M., Dougherty M.K.,
Khurana K.K., Russell, C.T.:
"Large-scale dynamics of {S}aturn's Magnetopause: Observations by Cassini",
{\it Journal of Geophysical Research 113}, A11209, 2008;
doi: 10.1029/2008JA013265


\bibitem[Andr\'e {\it et al.}(2008)]{andre-etal_2008}
Andr\'e N., Blanc M., Maurice S., Schippers P., Pallier E., Gombosi T.I.,
Hansen K.C., Young D.T., Crary F.J., Bolton S., Sittler E.C., Smith H.T.,
Johnson R.E., Baragiola R.A., Coates A.J., Rymer A.M., Dougherty M.K.,
Achilleos N., Arridge C.S., Krimigis S.M., Mitchell D.G., Krupp N.,
Hamilton D.C., Dandouras I., Gurnett D.A., Kurth W.S., Louarn P.,
Srama R., Kempf S., Waite H.J., Esposito L.W., Clarke J.T.:
"Identification of {S}aturn's Magnetospheric Regions and Associated Plasma Processes",
{\it Reviews of Geophysics 46}, RG4008, 2008;
doi: 10.1029/2007RG000238


\bibitem[Arridge {\it et al.}(2007)]{arridge-etal_2007}
Arridge C.S., Russell C.T., Khurana K.K., Achilleos N., 
Andr\'e N., Rymer A.M., Dougherty M.K., Coates A.J.:
"Mass of Saturn's magnetodisc: Cassini observations",
{\it Geophysical Research Letters 34}, L09108, 2007;
doi: 10.1029/2006GL028921


\bibitem[Arridge {\it et al.}(2009)]{arridge-etal_2009}
Arridge C.S., McAndrews H.J., Jackman C.M., Forsyth C., 
Walsh A.P., Sittler E.C., Gilbert L.K., Lewis G.R., Russell C.T.,
Coates A.J., Dougherty M.K., Collinson G.A., Wellbrock A.,
Young D.T.:
"Plasma electrons in Saturn's magnetotail: Structure, distribution and energisation",
{\it Planetary and Space Science 57}, 2032--2047, 2009;
doi: 10.1016/j.pss.2009.09.007


\bibitem[Bagenal \& Delamere(2011)]{bagenal-delamere_2011}
Bagenal F. \& Delamere P.A. (2011):
"Flow of mass and energy in the magnetospheres of Jupiter and Saturn",
{\it Journal of Geophysical Research 116}, A05209, 2011;
doi: 10.1029/2010JA016294


\bibitem[Banaszkiewicz \& Krivov(1997)]{bana-krivov_1997}
Banaszkiewicz M. \& Krivov A.V.:
"Hyperion as a Dust Source in the Saturnian System",
{\it Icarus 129}, p289--303, 1997;
doi: 10.1006/icar.1997.5781


\bibitem[Bertucci {\it et al.}(2011)]{bertucci-etal_2011}
Bertucci C., Duru F., Edberg N., Fraenz M., 
Martinecz C., Szego K., Vaisberg O.:
"The Induced Magnetospheres of Mars, Venus, and Titan",
{\it Space Science Reviews 162}, p113--171, 2011;
doi: 10.1007/s11214-011-9845-1


\bibitem[Bridge {\it et al.}(1981)]{bridge-etal_1981}
Bridge H.S., Belcher J.W., Lazarus A.J., Olbert S.,
Sullivan J.D., Bagenal F., Gazis P.R., Hartle R.E., Ogilvie K.W.,
Scudder J.D., Sittler E.C., Eviatar A., Siscoe G.L., Goertz C.K.,
Vasyliunas V.M.:
"Plasma Observations Near Saturn: Initial Results from Voyager 1",
{\it Science 212}, p217--224, 1981;
doi: 10.1126/science.212.4491.217


\bibitem[Bridge {\it et al.}(1982)]{bridge-etal_1982}
Bridge H.S., Bagenal F., Belcher J.W., Lazarus A.J.,
McNutt R.L., Sullivan J.D., Gazis P.R., Hartle R.E., Ogilvie K.W.,
Scudder J.D., Sittler E.C., Eviatar A., Siscoe G.L., Goertz C.K.,
Vasyliunas V.M.:
"Plasma Observations Near Saturn: Initial Results from Voyager 2",
{\it Science 215}, p563--570, 1982;
doi: 10.1126/science.215.4532.563


\bibitem[Burch {\it et al.}(2007)]{burch-etal_2007}
Burch J.L., Goldstein J., Lewis W.S., Young D.T.,
Coates A.J., Dougherty M.K., Andr\'e N.:
"Tethys and Dione as Sources of outward-flowing plasma in Saturn's magnetosphere",
{\it Nature 447}, p833--835, 2007;
doi: 10.1038/nature05906


\bibitem[Delamere {\it et al.}(2015)]{delamere-etal_2015}
Delamere P.A., Otto A., Ma X., Bagenal F., Wilson R.J:
"Magnetic flux circulation in the rotationally diven giant magnetospheres",
{\it Journal of Geophysical Research: Space Physics 120}, p4229--4245;
doi: 10.1002/2015JA021036


\bibitem[Frank {\it et al.}(1980)]{frank-etal_1980}
Frank L.A., Burek B.G., Ackerson K.L., Wolfe J.H., Mihalov J.D.:
"Plasmas in Saturn's Magnetosphere",
{\it Journal of Geophysical Research 85}, No. A11, p5695--5708, 1980;
doi: 10.1029/JA085iA11p05695


\bibitem[Fuselier {\it et al.}(2014)]{fuselier-etal_2014}
Fuselier S.A., Frahm R., Lewis W.S., Masters A., Mukherjee J., Petrinec S.M.,
Sillanpaa I.J.:
"The location of magnetic reconnection at Saturn's magnetopause:
A comparison with Earth",
{\it Journal of Geophysical Research Space Physics 119}, p2563--2578, 2014;
doi: 10.1002/2013JA019684


\bibitem[Galopeau \& Lecacheux(2000)]{galopeau-lecacheux_2000}
Galopeau P.H.M. \& Lecacheux A.:
"Variations of Saturn's radio rotation period measured at kilometer wavelengths",
{\it Journal of Geophysical Research 105}, No. A6, p13089--13102, 2000;
doi: 10.1029/1999JA005089


\bibitem[Goertz(1983)]{goertz_1983}
Goertz C.K.:
"Detached Plasma in Saturn's Front Side Magnetosphere",
{\it Geophysical Research Letters 10}, p455--458, 1983;
doi: 10.1029/GL010i006p00455


\bibitem[Goldreich \& Farmer(2007)]{goldreich-farmer_2007}
Goldreich P. \& Farmer A.J.:
"Spontaneous axisymmetry breaking of the external magnetic field at Saturn",
{\it Journal of Geophysical Research 112}, A05225, 2007;
doi: 10.1029/2006JA012163




\bibitem[Gr\"un {\it et al.}(1992)]{gruen-etal_1992}
Gr\"un E., Goertz C.K., Morfill G.E., Havnes O.:
"Statistics of Saturn's Spokes",
{\it Icarus 99}, p191--201, 1992;
doi: 10.1016/0019-1035(92)90182-7


\bibitem[Gr\"un {\it et al.}(2004)]{gruen-etal_2004}
Gr\"un E., Dikarev V., Frisch P.C., Graps A.,
Kempf S., Kr\"uger H., Landgraf M., Moragas-Klostermeyer G., Srama R.:
``Dust in Interplanetary Space and in the Local Galactic Environment'',
{\it ASP Conference Series 309}, 2004, p245--264;
in: Astrophysics of Dust,
Edited by Witt A.N., Clayton G.C., and Draine B.T.,
ISBN 978-1-58381-244-0


\bibitem[Gurnett {\it et al.}(2007)]{gurnett-etal_2007}
Gurnett D.A., Persoon A.M., Kurth W.S., Groene J.B.,
Averkamp T.F., Dougherty M.K., Southwood D.J.:
"The Variable Rotation Period of the Inner Region of Saturn's Plasma Disk",
{\it Science 316}, 2007, p442;
doi: 10.1126/science.1138562


\bibitem[Hartle {\it et al.}(1982)]{hartle-etal_1982}
Hartle R.E., Sittler E.C., Ogilvie K.W., Scudder J.D.,
Lazarus A.J., Atreya S.K.:
"Titan's Ion Exosphere Observed From Voyager 1",
{\it Journal of Geophysical Research 87}, A3, p1383--1394, 1982;
doi: 10.1029/JA087iA03p01383


\bibitem[Hill(1976)]{hill_1976}
Hill T.W.:
"Interchange stability of a rapidly rotating magnetosphere",
{\it Planetary and Space Science 24}, p1151--1154, 1976;
doi: 10.1016/0032-0633(76)90152-5


\bibitem[Hill {\it et al.}(2008)]{hill-etal_2008}
Hill T.W., Thomsen M.F., Henderson M.G., Tokar R.L., 
Coates A.J., McAndrews H.J., Lewis G.R., Mitchell D.G.,
Jackman C.M., Russell C.T., Dougherty M.K., Crary F.J., Young D.T.:
"Plasmoids in Saturn's magnetotail",
{\it Journal of Geophysical Research 113}, A01214, 2008;
doi: 10.1029/2007JA012626


\bibitem[Jackman {\it et al.}(2014)]{jackman-etal_2014}
Jackman C.M., Slavin J.A., Kivelson M.G., Southwood D.J.,
Achilleos N., Thomsen M.F., DiBraccio G.A., Eastwood J.P.,
Freeman M.P., Dougherty M.K., Vogt M.F.:
"Saturn's dynamic magnetotail: A comprehensive magngetic field
and plasma survey of plasmoids",
{\it Journal of Geophysical Research 119}, 5465--5494, 2014;
doi: 10.1002/2013JA019388


\bibitem[Kennedy {\it et al.}(2011)]{kennedy-etal_2011}
Kennedy G.M., Wyatt M.C., Su K.Y.L., Stansberry J.A.:
"Searching for Saturn's dust swarm: limits on the size distribution
   of irregular satellites from km to micron sizes",
{\it Monthly Notices RAS 417}, p2281--2287, 2011;
doi: 10.1111/j.1365-2966.2011.19409.x


\bibitem[Khalisi {\it et al.}(2015)]{khalisi-etal_2015}
Khalisi E., Srama R., Gr\"un E.:
"Counter data of the Cosmic Dust Analyzer aboard the
    Cassini spacecraft and possible 'dust clouds' at Saturn",
{\it Advances in Space Research 55}, p303--310, 2015;
doi: 10.1016/j.asr.2014.09.002


\bibitem[Krivov \& Banaszkiewicz(2001)]{krivov-bana_2001}
Krivov A.V. \& Banaszkiewicz M.:
"Unusual origin, evolution and fate of icy ejecta from Hyperion",
{\it Planetary and Space Science 49}, p1265--1279, 2001;
doi: 10.1016/S0032-0633(01)00069-1


\bibitem[Kr\"uger {\it et al.}(2010)]{krueger-etal_2010}
Kr\"uger H., Bindschadler D., Dermott S.F., Graps A.L.,
Gr\"un  E., Gustafson B.A., Hamilton D.P., Hanner M.S., Hor\'anyi M.,
Kissel J., Linkert D., Linkert G., Mann I., McDonnell J.A.M.,
Moissl R., Morfill G.E., Polanskey C., Roy M., Schwehm G., Srama, R.:
"Galileo dust data from the jovian system: 2000 to 2003",
{\it Planetary and Space Science 58}, p965-993, 2010;
doi: 10.1016/j.pss.2010.03.003




\bibitem[Porco {\it et al.}(2006)]{porco-etal_2006}
Porco C.C., Helfenstein P., Thomas P.C., Ingersoll A.P.,
Wisdom J., West R., Neukum G., Denk T., Wagner R., Roatsch T.,
Kieffer S., Turtle E., McEwen A., Johnson T.V., Rathbun J.,
Veverka J., Wilson D., Perry J., Spitale J., Brahic A.,
Burns J.A., Del Genio A.D., Dones L., Murray C.D., Squyres S.:
"Cassini Observes the Active South Pole of Enceladus",
{\it Science 311}, p1393--1401, 2006;
doi: 10.1126/science.1123013


\bibitem[Ragot \& Kahler(2003)]{ragot-kahler_2003}
Ragot B.R. \& Kahler S.W.:
"Interactions of Dust Grains with Coronal Mass Ejections
   and Solar Cycle Variations of the F-Coronal Brightness",
{\it The Astrophysical Journal 594}, p1049--1059, 2003;
doi: 10.1086/377076


\bibitem[Sittler {\it et al.}(1983)]{sittler-etal_1983}
Sittler E.C. jr., Ogilvie K.W., Scudder J.D.:
"Survey of Low-Energy Plasma Electrons in Saturn's Magnetosphere: Voyagers 1 and 2",
{\it Journal of Geophysical Research 88}, p8847--8870, 1983;
doi: 10.1029/JA088iA11p08847


\bibitem[Szego {\it et al.}(2005)]{szego-etal_2005}
Szego K., Bebesi Z., Erdos G., Foldy L.,
Crary F., McComas D.J., Young D.T., Bolton S., Coates A.J.,
Rymer A.M., Hartle R.E., Sittler E.C., Reisenfeld D.,
Bethelier J.J., Johnson R.E., Smith H.T., Hill T.W.,
Vilppola J., Steinberg J., Andre N.:
"The global plasma environment of Titan as observed by Cassini Plasma Spectrometer
     during the first two close encounters with Titan",
{\it Geophysical Research Letters 32}, L20S05, 2005,
doi: 10.1029/2005GL022646



\end{thebibliography}

\end{document}